\documentclass[12pt]{article}
\usepackage{graphicx,epsfig,epstopdf}
\usepackage{amssymb}
\begin{document}
\begin{center}
{ \LARGE {\bf On A New Form of Darboux-B\"{a}cklund Transformation for DNLS Equation-Mixed and Rational Type Solutions.
\\\
 } }
\end{center}
\vskip 0pt
\begin{center}
{\it {\large $Arindam\hskip 2 pt Chakraborty$ \\
 {Department of Physics,Heritage Institute of Technology,Kolkata-700107,India}\\
  and\\
  $A. Roy \hskip 3 ptChowdhury
$\footnote {(Corresponding author)e-mail: asesh\_r@yahoo.com}  }\\
\it{High Energy Physics Division, Department of Physics ,
                             Jadavpur University,\\
                              Kolkata - 700032,
                                    India}  }
\end{center}

\vskip 20pt
\begin{center}
{\bf Abstract}
\end{center}

\par A new form of Darboux-B\"{a}cklund transformation and its higher order form is derived for Derivative Nonlinear Schrodinger Equation(DNLS). The new form arises due to the different form of Lax pair. It is observed that by a special choice of the eigenvalue of DB transformation one can generate a mixed form of solution(containing both algebraic and exponential dependence on (x, t) can be generated. On the other hand by adopting a new methodology due to Neugebauer et. al. it is found that purely rational solution can be constructed. The two different approach yields different class of solution and are compared.

 \par PACS Number(s):  05.45.Pq, 05.45.Ac, 05.45.-a.

\section{Introduction}
Darboux-B\"{a}cklund transformation(DBT) is known to be a very useful tool to construct new solution of various non-linear equations which are associated with a Lax pair$^{[1]}$. Of course there exist other approaches for the generation of multisolitary states. Most versatile of them is the Inverse Scattering transform$^{[2]}$. But that requires a detailed analyticity study of the spectral problem in the complex eigenvalue plain$^{[3]}$ for establishing the Gelfand-Levitan-Marchenko integro-differential equation$^{[4]}$. Such an analysis is quite elaborate and time consuming. That is why Darboux-B\"{a}cklund transformation is now a days so popular for the construction of a bigger class of solitary solutions$^{[5]}$. Also, in this approach one has the freedom of choice for the seed solutions. Here, we have adopted both the traditional approach$^{[6]}$, which usually uses two eigenfunction at a time and the Neugebuer method$^{[7]}$ for which it is sufficient to fix one eigenvalue and corresponding eigenfunction$^{[8]}$. These two methods lead to two different class of solutions, one is a mixture of rational and exponential function and the other is purely rational solution.
\section{Formulation}
There exist various forms of Lax pair for the Derivative Nonlinear Schrodinger Equation(DNLS),from which we consider the following one given in reference(9). The equations are
\begin{eqnarray}
q_t+iq_{xx}+(rq^2)_x=0\nonumber\\
r_t-ir_{xx}+(r^2q)_x=0
\end{eqnarray}
and the Lax pair is
\begin{eqnarray}
\Psi_x=M\Psi\nonumber\\
\Psi_t=N\Psi
\end{eqnarray}
with
\begin{eqnarray}
M=J\lambda^2-R\lambda\nonumber\\
N=2J\lambda^4-2R\lambda^3+qrJ\lambda^2+U\lambda
\end{eqnarray}
where
\begin{equation}
J=\left(\begin{array}{cc}
   i & 0 \\
   0 & -i
   \end{array} \right)\\
\end{equation}
\begin{equation}
R=\left(\begin{array}{cc}
   0 & q \\
   r & 0
   \end{array} \right)\\
\end{equation}
\begin{equation}
U=\left(\begin{array}{cc}
   0 & -iq_x-rq^2 \\
   ir_x-r^2q & 0
   \end{array} \right)\\
\end{equation}
The eigenfunction $\Psi$ is written as
\begin{equation}
\Psi=\left(\begin{array}{cc}
   \psi_{11} & \psi_{12} \\
   \psi_{21} & \psi_{22}
   \end{array} \right)\\
\end{equation}
If the seed solution is chosen as $q=q_0\exp[i(kx-\omega t)]$ and $r=r_0\exp[-i(kx-\omega t)]$ with $q_0$ and $r_0$ being constants and satisfying equation(1)
then the Lax eigenfunctions turn out to be
\begin{eqnarray}
\psi_{11}(x, t)&=&\frac{i\lambda q_0}{\alpha_{+}-\lambda^2}\exp[i(\alpha_{+}x-\beta_{+}t)]\nonumber\\
\psi_{21}(x, t)&=&-\frac{i(P(\lambda)+\beta_{+})}{Q(\lambda)}\exp[i((\alpha_{+}-k)x-(\beta_{+}-\omega)t)]\nonumber\\
\psi_{12}(x, t)&=&\frac{i\lambda q_0}{\alpha_{-}-\lambda^2}\exp[i(\alpha_{-}x-\beta_{-}t)]\nonumber\\
\psi_{22}(x, t)&=&-\frac{i(P(\lambda)+\beta_{-})}{Q(\lambda)}\exp[i((\alpha_{-}-k)x-(\beta_{-}-\omega)t)]
\end{eqnarray}
with
\begin{eqnarray}
P(\lambda)&=&2\lambda^4+\lambda^2q_0r_0\nonumber\\
Q(\lambda)&=&-\lambda r_0q_0^2+\lambda k q_0-2\lambda^3q_0\nonumber\\
R(\lambda)&=&-2\lambda^3 r_0+\lambda k r_0-\lambda r_0^2 q_0
\end{eqnarray}
\begin{eqnarray}
\alpha_{\pm}=\frac{1}{2}[k\pm \{k^2+4\lambda^2(\lambda^2-k-q_0r_0)\}^{1/2}]\nonumber\\
\beta_{\pm}=\frac{1}{2}[\omega\pm \{\omega^2+4(P^2+\omega P-R Q)\}^{1/2}]
\end{eqnarray}
\par Next we choose the Darboux-B\"{a}cklund transformation
\begin{equation}
\Psi^{\prime}=D\Psi
\end{equation}
where, $\Psi^{\prime}$ satisfies the similar Lax equations as (2) with new nonlinear fields $(q^{\prime}, r^{\prime})$. In our case $D(\lambda)$
\begin{equation}
D(\lambda)=\lambda^2\left(\begin{array}{cc}
   a_2 & 0 \\
   0 & d_2
   \end{array} \right)\\
   +\lambda\left(\begin{array}{cc}
   0 & b_1 \\
   c_1 & 0
   \end{array} \right)\\+
   \left(\begin{array}{cc}
   a_0 & 0 \\
   0 & d_0
   \end{array} \right)\\
\end{equation}
The equation satisfied by $D(\lambda)$ is
\begin{eqnarray}
D_x=M^{\prime}D-D M\nonumber\\
D_t=N^{\prime}D-D N
\end{eqnarray}
It may be added at this point that the usual choice of $D$
\begin{equation}
D(\lambda)=
   \lambda\left(\begin{array}{cc}
   a_1 & b_1 \\
   c_1 & d_1
   \end{array} \right)\\+
   \left(\begin{array}{cc}
   a_0 & b_0 \\
   c_0 & d_0
   \end{array} \right)\\
\end{equation}
does not hold in our case because the equation satisfied by $D$(equation(13)) leads to a trivial solution $D=$null matrix. This is happening due to the form of the Lax pair(equation(2))which does not contain any term free from $\lambda$.
\par From the condition that $\det D(\lambda)=0$ at $\lambda=\lambda_1$ and $\lambda_2$, we can determine
\begin{eqnarray}
a_2&=&\frac{a_0(\lambda_1\phi_1\psi_2-\lambda_2\psi_1\phi_2)}{\lambda_1\lambda_2(\lambda_1\psi_1\phi_2-\lambda_2\psi_2\phi_1)}\nonumber\\
d_2&=&\frac{d_0(\lambda_2\phi_1\psi_2-\lambda_1\psi_1\phi_2)}{\lambda_1\lambda_2(\lambda_2\psi_1\phi_2-\lambda_1\psi_2\phi_1)}\nonumber\\
b_1&=&\frac{a_0(\lambda_2^2-\lambda_1^2)\psi_1\phi_1}{\lambda_1\lambda_2(\lambda_1\psi_1\phi_2-\lambda_2\psi_2\phi_1)}\nonumber\\
c_1&=&-\frac{d_0(\lambda_2^2-\lambda_1^2)\psi_2\phi_2}{\lambda_1\lambda_2(\lambda_2\psi_1\phi_2-\lambda_1\psi_2\phi_1)}
\end{eqnarray}
along with the relations
\begin{eqnarray}
q^{\prime}=a_2^2 q+2ib_1a_2\nonumber\\
r^{\prime}=d_2^2r-2ic_1d_2
\end{eqnarray}
If one now uses eigenfunctions(8) in these one can construct a pair of nontrivial solutions, by starting with a constant solution. These can be finally written as
\begin{eqnarray}
q^{\prime}=\frac{\Lambda^2(\lambda_2,\lambda_1)}{\Lambda^2(\lambda_1,\lambda_2)}\left(q+2i\frac{\lambda_1^2-\lambda_2^2}{\Lambda(\lambda_2,\lambda_1)}\psi_1\phi_1\right)\nonumber\\
r^{\prime}=\frac{\Lambda^2(\lambda_1,\lambda_2)}{\Lambda^2(\lambda_2,\lambda_1)}\left(r+2i\frac{\lambda_1^2-\lambda_2^2}{\Lambda(\lambda_1,\lambda_2)}\psi_2\phi_2\right)
\end{eqnarray}
where
\begin{eqnarray}
\Lambda(\lambda_2, \lambda_1)&=&\lambda_2\psi_1\phi_2-\lambda_1\psi_2\phi_1\nonumber\\
&=&\lambda_1\lambda_2q_0 \Theta
\end{eqnarray}
where
\begin{eqnarray}
\Theta=\left(\frac{P(\lambda_2)+\sigma_2}{(\mu_1-\lambda_1^2)Q(\lambda_2)}-\frac{P(\lambda_1)+\sigma_1}{(\mu_2-\lambda_2^2)Q(\lambda_1)}\right)\exp (i\bigsqcup (\mu_1, \mu_2, \sigma_1, \sigma_2, k, \omega, x, t))\nonumber\\
+k_2\left(\frac{P(\lambda_2)+\delta_2}{(\mu_1-\lambda_1^2)Q(\lambda_2)}-\frac{P(\lambda_1)+\sigma_1}{(\nu_2-\lambda_2^2)Q(\lambda_1)}\right)\exp(i\bigsqcup (\mu_1, \nu_2, \sigma_1, \delta_2, k, \omega, x, t))\nonumber\\
+k_1\left(\frac{P(\lambda_2)+\sigma_2}{(\nu_1-\lambda_1^2)Q(\lambda_2)}-\frac{P(\lambda_1)+\delta_1}{(\mu_2-\lambda_2^2)Q(\lambda_1)}\right)\exp(i\bigsqcup (\nu_1, \mu_2, \delta_1, \sigma_2, k, \omega, x, t))\nonumber\\
+k_1k_2\left(\frac{P(\lambda_2)+\delta_2}{(\nu_1-\lambda_1^2)Q(\lambda_2)}-\frac{P(\lambda_1)+\delta_1}{(\nu_2-\lambda_2^2)Q(\lambda_1)}\right)\exp(i\bigsqcup (\nu_1, \nu_2, \delta_1, \delta_2, k, \omega, x, t))
\end{eqnarray}
Here, the function $\bigsqcup$ is given by
\begin{eqnarray}
\bigsqcup(a, b, c, d, k,\omega, x, t )=(a+b-k)x-i(c+d-\omega)t
\end{eqnarray}
with a similar expressions for other $\Lambda's$.
\section{Second Order Transformation}
One of the most important properties of  Darboux-B\"{a}cklund transformation it can be repeated for finite number of times to generate higher solutions. Let us consider that the second time transformed eigenfunction is $\Psi^{\prime\prime}$;$\Psi^{\prime\prime}=D_1\Psi^{\prime}=D_1D\Psi$. Here, $D_1$ stands for the next DB operator written as
\begin{equation}
D_1(\lambda)=\lambda^2\left(\begin{array}{cc}
   A_2 & 0 \\
   0 & D_2
   \end{array} \right)\\
   +\lambda\left(\begin{array}{cc}
   0 & B_1 \\
   C_1 & 0
   \end{array} \right)\\+
   \left(\begin{array}{cc}
   A_0 & 0 \\
   0 & D_0
   \end{array} \right)\\
\end{equation}
As per previous formulae
\begin{eqnarray}
A_2&=&\frac{A_0(\lambda_1\phi_1^{\prime}\psi_2^{\prime}-\lambda_2\psi_1^{\prime}\phi_2^{\prime})}{\lambda_1\lambda_2(\lambda_1\psi_1^{\prime}\phi_2^{\prime}-\lambda_2\psi_2^{\prime}\phi_1^{\prime})}\nonumber\\
D_2&=&\frac{D_0(\lambda_2\phi_1^{\prime}\psi_2^{\prime}-\lambda_1\psi_1^{\prime}\phi_2^{\prime})}{\lambda_1\lambda_2(\lambda_2\psi_1^{\prime}\phi_2^{\prime}-\lambda_1\psi_2\phi_1)}\nonumber\\
B_1&=&\frac{A_0(\lambda_2^2-\lambda_1^2)\psi_1^{\prime}\phi_1^{\prime}}{\lambda_1\lambda_2(\lambda_1\psi_1^{\prime}\phi_2^{\prime}-\lambda_2\psi_2^{\prime}\phi_1^{\prime})}\nonumber\\
C_1&=&-\frac{D_0(\lambda_2^2-\lambda_1^2)\psi_2^{\prime}\phi_2^{\prime}}{\lambda_1\lambda_2(\lambda_2\psi_1^{\prime}\phi_2^{\prime}-\lambda_1\psi_2^{\prime}\phi_1^{\prime})}
\end{eqnarray}
where, $\Psi^{\prime}=D\Psi$. After a lot of simplification we get
\begin{eqnarray}
q^{\prime\prime}=\frac{\Lambda^{\prime 2}(\lambda_2,\lambda_1)}{\Lambda^{\prime 2}(\lambda_1,\lambda_2)}\left(q^{\prime}+2i\frac{\lambda_1^2-\lambda_2^2}{\Lambda^{\prime}(\lambda_2,\lambda_1)}\psi^{\prime}_1\phi^{\prime}_1\right)\nonumber\\
r^{\prime\prime}=\frac{\Lambda^{\prime 2}(\lambda_1,\lambda_2)}{\Lambda^{\prime 2}(\lambda_2,\lambda_1)}\left(r^{\prime}+2i\frac{\lambda_1^2-\lambda_2^2}{\Lambda^{\prime}(\lambda_1,\lambda_2)}\psi^{\prime}_2\phi^{\prime}_2\right)
\end{eqnarray}
\section{Mixed Rational Solution}
\subsection{Case 1}
Now let us start from a set of constant seed solutions i.e.; $q_0=const$ and $r_0=const$ and an eigenvalue $\lambda=\sqrt{q_0r_0}$. Then it is easy to ascertain that the Lax eigenfunctions are
\begin{eqnarray}
\psi_{1}(\lambda_1)&=&\psi_{11}(\lambda_1)+k_1 \psi_{12}(\lambda_1)\nonumber\\
&=&(A+k_1A^{\prime})xt\nonumber\\
\psi_{2}(\lambda_1)&=&\psi_{21}(\lambda_1)+k_1 \psi_{22}(\lambda_1)\nonumber\\
&=&(C+k_1C^{\prime})x+(D+k_1D^{\prime})t
\end{eqnarray}
Let us choose $\lambda_2=\lambda_0$ any fixed constant eigenvalue and then the second set of eigenfunction turns out to be
\begin{eqnarray}
\psi_{11}(\lambda_0)=\exp[\zeta_{+}x+\delta_{+}t]\nonumber\\
\psi_{21}(\lambda_0)=\exp[\zeta_{-}x+\delta_{-}t]
\end{eqnarray}
whence;
\begin{eqnarray}
\phi_1(\lambda_2)&=&\phi_1(\lambda_0)\nonumber\\
&=&\psi_{11}(\lambda_0)+k_2\psi_{12}(\lambda_0)\nonumber\\
&=&(1+k_2)\exp[\zeta_{+}x+\delta_{+}t]\nonumber\\
\phi_1(\lambda_2)&=&(1+k_2)\exp[\zeta_{-}x+\delta_{-}t]
\end{eqnarray}
where, $\delta_{\pm}$ are solutions of
\begin{equation}
\delta^2+(2\lambda^4_0+\lambda^2_0q_0r_0)^2-(2q_0\lambda^3_0+r_0q_0^2)(2\lambda_0^3r_0-r_0^2q_0)=0
\end{equation}
Now if these two sets of eigenfunctions are used in formulae(17) we get
\begin{eqnarray}
q^{\prime}=q_0\left[\frac{\mu_1xt\exp(\zeta_{-}x+\delta_{-}t)-(\mu_2x+\mu_3t)\exp(\zeta_{+}x+\delta_{+}t)}{(\mu_4x+\mu_5t)\exp(\zeta_{+}x+\delta_{+}t)-\mu_6xt\exp(\zeta_{-}x+\delta_{-}t)}\right]\nonumber\\
+2i\left[\frac{\mu_7xt\exp(\zeta_{-}x+\delta_{-}t)-(\mu_8x+\mu_9t)\exp(\zeta_{+}x+\delta_{+}t)}{(\mu_4x+\mu_5t)\exp(\zeta_{+}x+\delta_{+}t)-\mu_6xt\exp(\zeta_{-}x+\delta_{-}t)}\right]\nonumber\\
\times (\lambda_0^2-q_0r_0)\mu_{10}xt\exp(\zeta_{+}x+\delta_{+}t)
\end{eqnarray}

\begin{figure}
\begin{center}
\includegraphics[scale=0.5,angle=0]{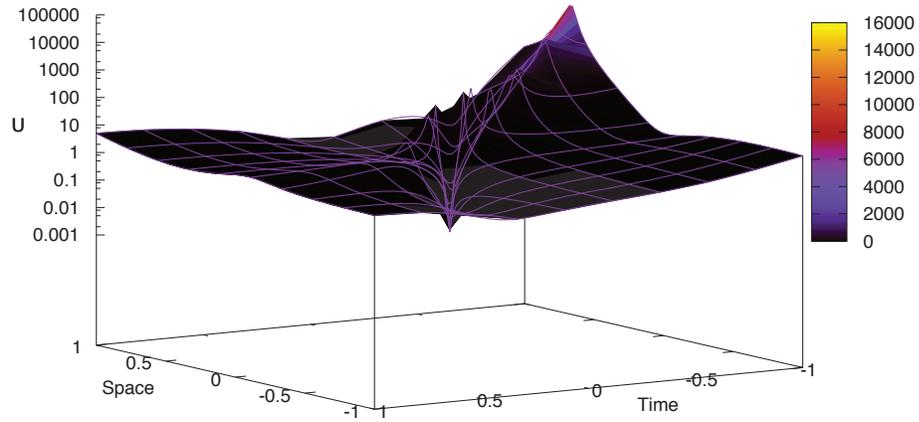}
\end{center}
\caption{U vs spacetme}
\label{fig:1}
\end{figure}

\begin{figure}
\begin{center}
\includegraphics[scale=0.8]{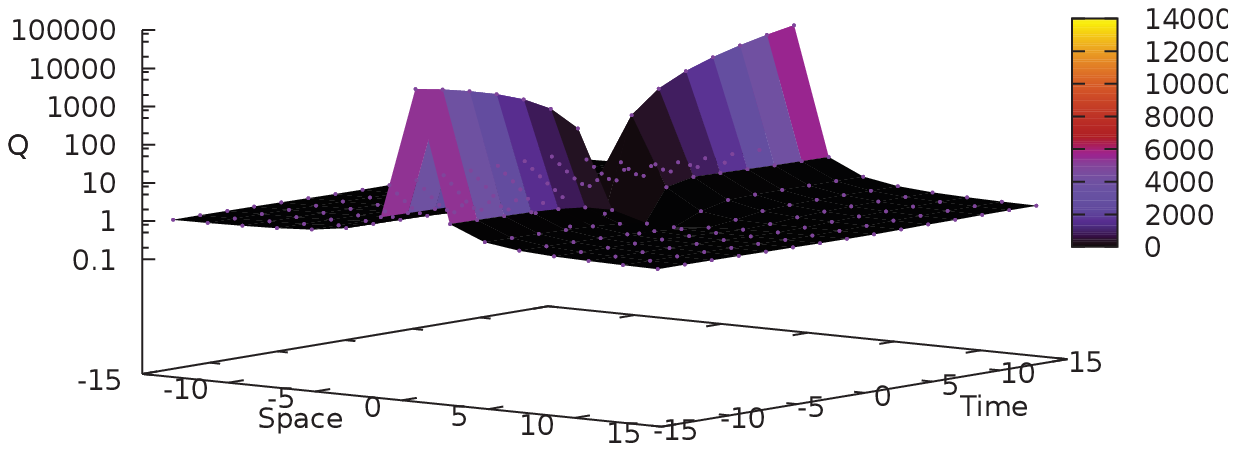}
\end{center}
\caption{a}
\label{fig:2}
\end{figure}
\begin{figure}
\begin{center}
\includegraphics[scale=0.5]{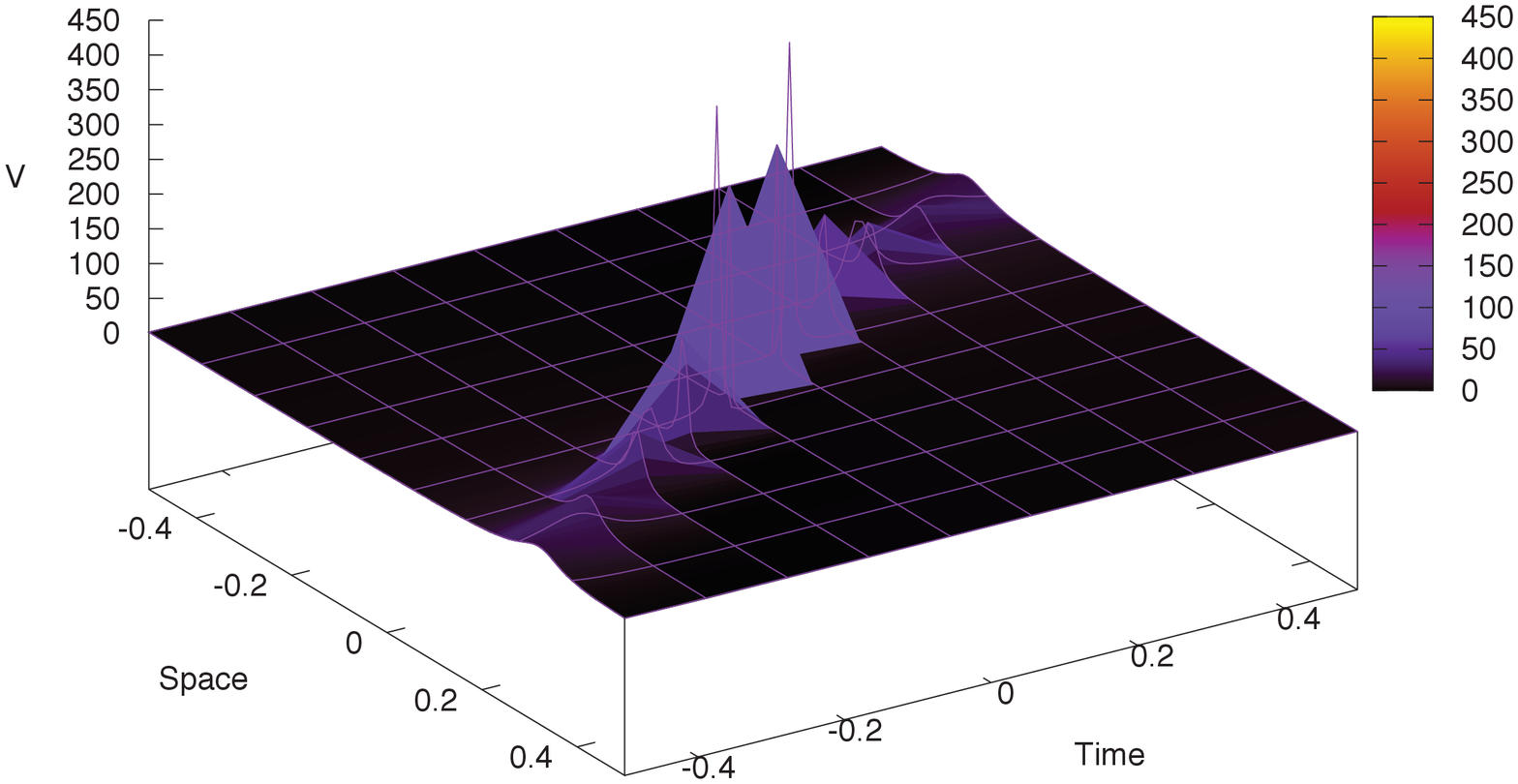}
\end{center}
\caption{a}
\label{fig:2}
\end{figure}

Where, $\mu_i$ are constants being combination of those occurring in the expressions of $\phi_i$, $\psi_i$'s. The second formulae of (17) can be used similarly to construct an expression for $r^{\prime}$ also. It is interesting to see that the above expression is a mixture of algebraic expressions in $(x, t)$ and also exponential functions.
Fig()represents $U=\mid q^{\prime} \mid^2$ for $\mu_1=-\mu_6=\mu_{10}=\mu_7=q_0=r_0=1$, $\mu_2=\mu_5=\mu_8=2i$ and $\mu_3=\mu_4=\mu_9=3i$ with $\zeta_{+}=\zeta_{-}=\delta_{-}=-\delta_{+}=1$ and $\lambda_0=\sqrt 2$. Fig() represents the $V=\mid q^{\prime} \mid^2$ choosing  $\delta_{+}=1$ retaining all the other values of  the constants unchanged.
\subsection{Case 2}
On the other hand a pair of nonconstant seed solutions of the following form
\begin{eqnarray}
q=q_0\exp(-iq_0^2x)\nonumber\\
r=-q_0\exp(iq_0^2x)
\end{eqnarray}
leads us to mixed-rational solutions for eigenvalues $\lambda=q_0\frac{i\pm 1}{2}$.Following the same procedure as described above we obtain the 1st order transformed solution of Darboux-B\"{a}cklund type
\begin{eqnarray}
q^{\prime}=q_0\exp(-iq_0^2x)\left[\frac{1+2iq_0^2x-2q_0^4(1-q_0^4)t-q_0^4x^2-2iq_0^6xt}{1-2iq_0^2x+2q_0^4(1-q_0^4)t+q_0^4x^2+2iq_0^6xt}\right]^2\nonumber\\
\times\left[1-2\frac{1+2q_0^4(1+q_0^4)t-q_0^4x^2-2iq_0^6xt}{1+2iq_0^2x-2q_0^4(1-q_0^4)t-q_0^4x^2-2iq_0xt}\right]
\end{eqnarray}
and
\begin{eqnarray}
r^{\prime}=-q_0\exp(iq_0^2x)\left[\frac{1-2iq_0^2x+2q_0^4(1-q_0^4)t+q_0^4x^2+2iq_0^6xt}{1+2iq_0^2x-2q_0^4(1-q_0^4)t-q_0^4x^2-2iq_0^6xt}\right]^2\nonumber\\
\times\left[1+2\frac{1-2q_0^4(1-q_0^4)t-q_0^4x^2-2iq_0^6xt}{1-2iq_0^2x+2q_0^4(1-q_0^4)t+q_0^4x^2+2iq_0^6xt}\right]
\end{eqnarray}
Fig() and fig() represent $Q=\mid q^{\prime}\mid^2$ and $R=\mid r^{\prime}\mid^2$ with $q_0=1$. Fig() and fig() represent $Q$ vs $x$ keeping $t$ fixed and $Q$ vs $t$ keeping $x$ fixed while fig() and fig() represent the corresponding diagrams for $R$.

\begin{figure}
\begin{center}
\includegraphics[scale=0.8]{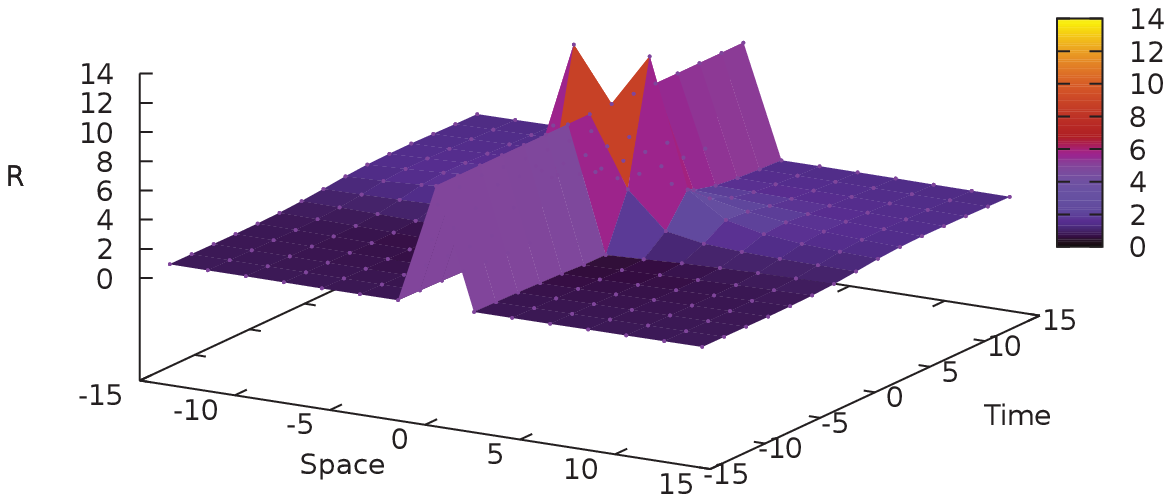}
\end{center}
\caption{a}
\label{fig:2}
\end{figure}

\begin{figure}
\begin{center}
\includegraphics[scale=0.7]{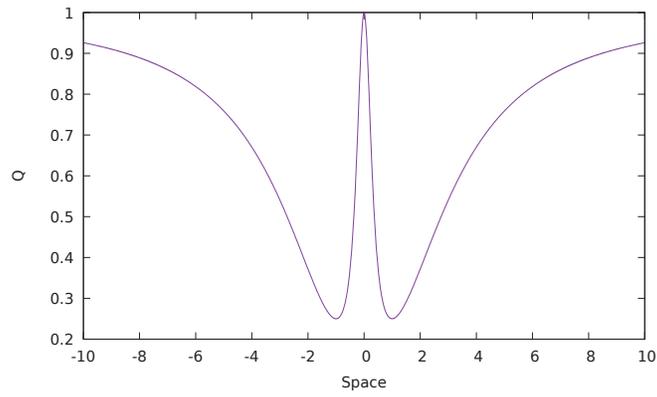}
\end{center}
\caption{Space vs Q at time}
\end{figure}
\begin{figure}
\begin{center}
\includegraphics[scale=0.7]{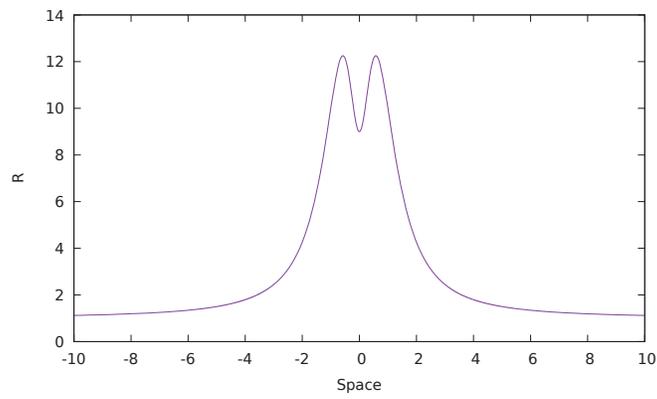}
\end{center}
\caption{Space vs R at time}
\end{figure}
\begin{figure}
\begin{center}
\includegraphics[scale=0.7]{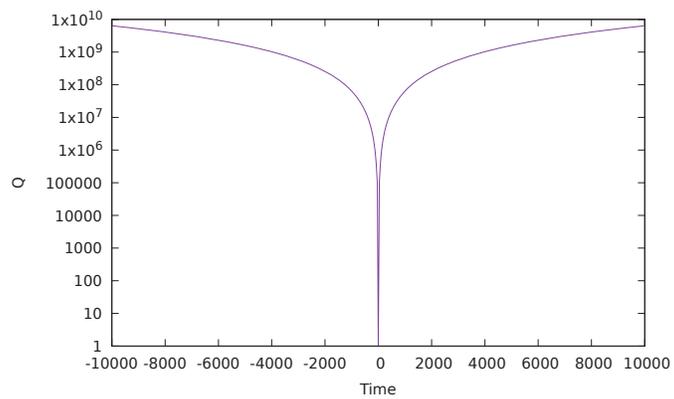}
\end{center}
\caption{two dimensional plot}
\end{figure}
\begin{figure}
\begin{center}
\includegraphics[scale=0.7]{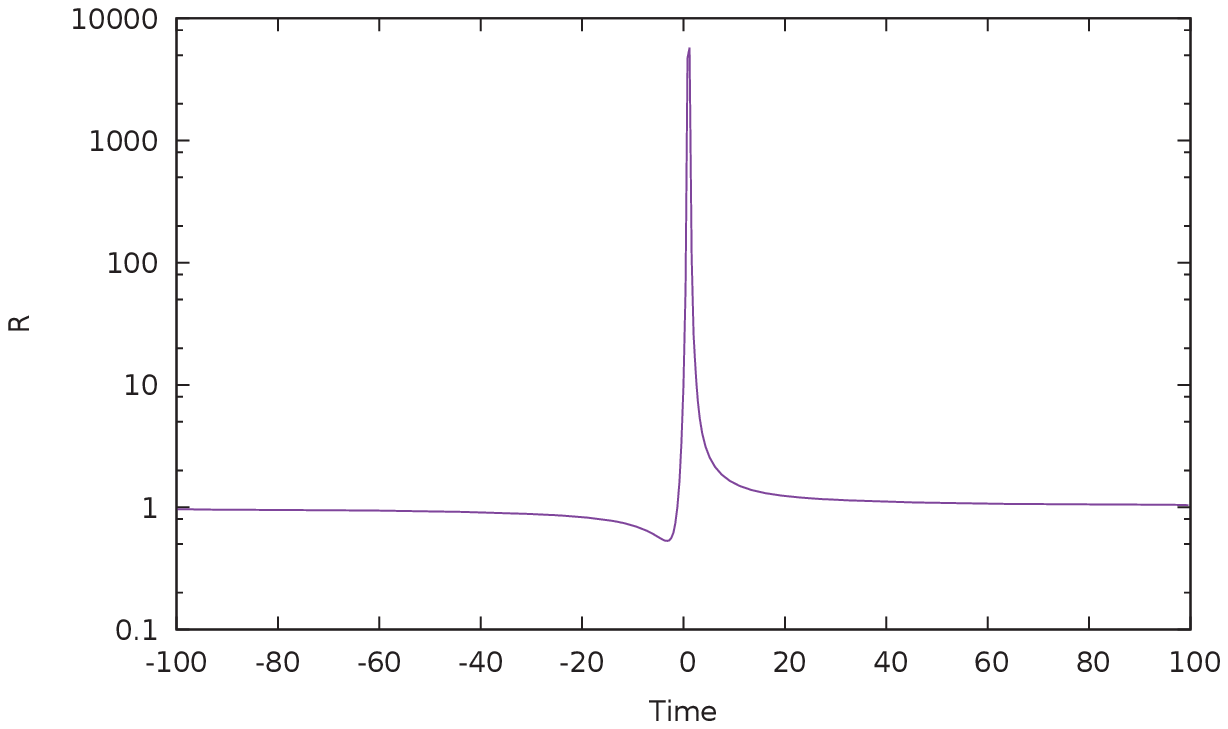}
\end{center}
\caption{projection on 2d}
\end{figure}


\section{Comments and Discussion}
In our above analysis we have shown that by starting with two constant solutions(case-1) or non-constant seed solutions(case-2) it is possible to generate anew form of mixed rational solutions (for the DNLS equation) which contain both polynomial in $(x, t)$ as well as exponential functions of the same. This was accomplished with help of Darboux-B\"{a}cklund transformation and a special choice of $(\lambda_1, \lambda_2)$. On the other hand a different approach to DBT was proposed by Neugebauer et. al. which does not require two such eigenfunctions and eigenvalues. In this approach we write our Lax operator as
\begin{eqnarray}
L=\lambda^2 L_2-\lambda L_1
\end{eqnarray}
and the starting one as $L_0$, then DBT matrix say $P$ should satisfy
\begin{eqnarray}
LP=P_x+P L_0
\end{eqnarray}
where $P$ is written as
\begin{equation}
P(x,t,\lambda)=\sum_{j=1}^{N-1}\lambda^jP_j+\lambda^N \mathbf{1}
\end{equation}
with
\begin{equation}
P_j=\left(\begin{array}{cc}
   A_j & B_j \\
   C_j & D_j
   \end{array} \right)
\end{equation}
whence equation(33)written in full can be seen to be equal to
\begin{eqnarray}
\left[i\lambda^2
\left(\begin{array}{cc}
   1 & 0 \\
   0 & -1
   \end{array} \right)
-\lambda
\left(\begin{array}{cc}
   0 & q \\
   r & 0
   \end{array} \right)\right]\left[\sum_{j=1}^{N-1}\lambda^j
\left(\begin{array}{cc}
   A_j & B_j \\
   C_j & D_j
   \end{array} \right)
+\lambda^N
\left(\begin{array}{cc}
   1 & 0 \\
   0 & 1
   \end{array} \right)\right]\nonumber\\
=\sum_{j=1}^{N-1}\lambda^j
\left(\begin{array}{cc}
   A_{jx} & B_{jx} \\
   C_{jx} & D_{jx}
   \end{array} \right)\\
   +
\left[\sum_{j=1}^{N-1}\lambda^j
\left(\begin{array}{cc}
   A_{j} & B_{j} \\
   C_{j} & D_{j}
   \end{array} \right)+\lambda^N
\left(\begin{array}{cc}
   1 & 0 \\
   0 & 1
   \end{array} \right)\right]\left[i\lambda^2
\left(\begin{array}{cc}
   1 & 0 \\
   0 & -1
   \end{array} \right)
-\lambda
\left(\begin{array}{cc}
   0 & q_0 \\
   r_0 & 0
   \end{array} \right)\right]
\end{eqnarray}
Equating the coefficients of $\lambda^{N+1}$ we get

\begin{eqnarray}
q=q_0+2iB_{N-1}\nonumber\\
r=r_0-2iC_{N-1}
\end{eqnarray}
where as the coefficients $B_N$, $C_N$ etc. are obtained as solutions of
\begin{eqnarray}
\sum_{j=1}^{N-1}(A_j+\beta_i B_j)\lambda^j_i=-\lambda^N_i\nonumber\\
\sum_{j=1}^{N-1}(D_j+\beta^{-1}_i C_j)\lambda^j_i=-\lambda^N_i
\end{eqnarray}
with $(\lambda_i\mid i=1, 2, \cdots 2N)$ being $2N$ solutions of $\det P$.
Solving equation()by Cramer's  rule leads to
\begin{eqnarray}
B_{N-1}=\frac{\Delta^{(B)}}{\Delta_1}\nonumber\\
C_{N-1}=\frac{\Delta^{(C)}}{\Delta_2}
\end{eqnarray}
where $\Delta's$ are determinants written below
\begin{eqnarray}
\Delta_1= \left|\begin{array}{cccccccccc}
   1 & \beta_1 &\lambda_1&\beta_1\lambda_1&\lambda_1^2&\ldots&\lambda_1^{N-1}&\lambda_1^{N-1}\beta_1\\
   1 & \beta_2&\lambda_2&\beta_2\lambda_2&\lambda_2^2&\ldots&\lambda_2^{N-1}&\lambda_2^{N-1}\beta_2\\
   \vdots&\vdots&\ddots&\vdots\\
   1&\beta_{2N}&\lambda_{2N}&\beta_{2N}\lambda_{2N}&\lambda_{2N}^2&\ldots&\lambda_{2N}^{N-1}&\lambda_{2N}^{N-1}\beta_{2N}
 \end{array} \right|
\end{eqnarray}
\begin{eqnarray}
\Delta^{(B)}= \left|\begin{array}{cccccccccc}
   1 & \beta_1 &\lambda_1&\beta_1\lambda_1&\lambda_1^2&\ldots&\lambda_1^{N-1}&\lambda_1^{N}\\
   1 & \beta_2&\lambda_2&\beta_2\lambda_2&\lambda_2^2&\ldots&\lambda_2^{N-1}&\lambda_2^{N}\\
   \vdots&\vdots&\ddots&\vdots\\
   1&\beta_{2N}&\lambda_{2N}&\beta_{2N}\lambda_{2N}&\lambda_{2N}^2&\ldots&\lambda_{2N}^{N-1}&\lambda_{2N}^{N}
 \end{array} \right|
\end{eqnarray}
\begin{eqnarray}
\Delta_2= \left|\begin{array}{cccccccccc}
   1 & \beta^{-1}_1 &\lambda_1&\beta^{-1}_1\lambda_1&\lambda_1^2&\ldots&\lambda_1^{N-1}&\lambda_1^{N-1}\beta^{-1}_1\\
   1 & \beta^{-1}_2&\lambda_2&\beta^{-1}_2\lambda_2&\lambda_2^2&\ldots&\lambda_2^{N-1}&\lambda_2^{N-1}\beta^{-1}_2\\
   \vdots&\vdots&\ddots&\vdots\\
   1&\beta^{-1}_{2N}&\lambda_{2N}&\beta^{-1}_{2N}\lambda_{2N}&\lambda_{2N}^2&\ldots&\lambda_{2N}^{N-1}&\lambda_{2N}^{N-1}\beta^{-1}_{2N}
 \end{array} \right|
\end{eqnarray}
and
\begin{eqnarray}
\Delta^{(C)}= \left|\begin{array}{cccccccccc}
   1 & \beta^{-1}_1 &\lambda_1&\beta^{-1}_1\lambda_1&\lambda_1^2&\ldots&\lambda_1^{N-1}&\lambda_1^{N}\\
   1 & \beta^{-1}_2&\lambda_2&\beta^{-1}_2\lambda_2&\lambda_2^2&\ldots&\lambda_2^{N-1}&\lambda_2^{N}\\
   \vdots&\vdots&\ddots&\vdots\\
   1&\beta^{-1}_{2N}&\lambda_{2N}&\beta^{-1}_{2N}\lambda_{2N}&\lambda_{2N}^2&\ldots&\lambda_{2N}^{N-1}&\lambda_{2N}^{N}
 \end{array} \right|
\end{eqnarray}
where $\beta_i$ stands for
\begin{equation}
\beta_i=\frac{\psi_{21}^0-b_i\psi_{22}^0}{\psi_{11}^0-b_i\psi_{12}^0}
\end{equation}
Suppose $\lambda=\sqrt{q_0r_0}$ and $P(\lambda)$ is so chosen that it has a root at $\lambda_1$, then
\begin{equation}
\beta_i=\beta_i(\lambda_1)=\frac{\psi_{21}^0(\lambda_1)-b_i(\lambda_1)\psi_{22}^0(\lambda_1)}{\psi_{11}^0(\lambda_1)-b_i(\lambda)\psi_{12}^0(\lambda_1)}
\end{equation}
Choose as before $\psi_{21}^0=Axt$, $\psi_{22}^0=A^{\prime}xt$, $\psi_{11}^0=Cx+Dt$ and $\psi_{12}^0=C^{\prime}x+D^{\prime}t$,
whence
\begin{equation}
\beta_1=\frac{(A-b_1A^{\prime})xt}{(C-b_1C^{\prime})x+(D-b_1D^{\prime})t}
\end{equation}
 the whole form of the solution equation(38)is composed of $\beta_1$ and its powers. So it is now purely rational one.
In a similar vein for nonconstant seed solutions as taken in equation(29) one can construct $\beta_1$ corresponding to both $\lambda_{\pm}=q_0\frac{i\pm 1}{2}$ with respective seed eigenfunctions given below.
\par For $\lambda=\lambda_{+}$
\begin{eqnarray}
\psi^0_{11}(\lambda_{+})=A_0[1-q_0^2(x+q_0^2(i-1)t)]\exp(-iq_0^2x/2)\nonumber\\
\psi^0_{21}(\lambda_{+})=B_0[1+q_0^2(x+q_0^2(i-1)t)]\exp(iq_0^2x/2)\nonumber\\
\end{eqnarray}

For $\lambda=\lambda_{-}$
\begin{eqnarray}
\psi^0_{11}(\lambda_{-})=A^{\prime}_0[1+q_0^2(x+q_0^2(i-1)t)]\exp(-iq_0^2x/2)\nonumber\\
\psi^0_{21}(\lambda_{-})=B^{\prime}_0[1-q_0^2(x+q_0^2(i-1)t)]\exp(iq_0^2x/2)\nonumber\\
\end{eqnarray}
Assuming similar expressions for $\psi_{12}$ and $\psi_{22}$ corresponding $\beta_1$ can be constructed involving both rational and exponential part. this when put in the expressions of $q^{\prime}$ and $r^{\prime}$ gives mixed-rational solutions.
So, with the help of Nuegebauer approach we have been able to construct both purely rational and mixed solutions of DNLS equation. The basic difference between these two methodologies is that in the Neugebauer approach we need one eigenfunction and one eigenvalue in contrast to the traditional one. So it can be summarized to DBT can be effectively utilized to generate new class of solutions of integrable nonlinear systems.

\section{references}

\par [1] B\"{a}cklund Transformation-Eds. R. Eckmann and R. K. Dodd in Series Lecture notes in Mathematics Vol\textbf{505} Springer-Verlag Berlin 1980\\

\par [2] Ablowitz, Kaup, Newell, Segur- Stud. App. Math.\\

\par [3]M. Ablowitz and A. Fokas-Complex Variable(Cambridge)\\

\par [4] Theory of Solitons Consultants Bureau New York, London L S. Novikov, S. V. Manakov, L. P. Pitaevskii and V. E. Zakharov\\

\par [5] M. A. Salle and V. B. Mateev-Darboux transformation and solitons-Springer-Verlag Berlin(1991).\\

\par [6] S. Tian, Tian-tian Zhang, H. quing Zhang-Phys. Scr. \textbf{80}(2009)065013.\\

\par [7]  G. Neugebauer, R. Meiner-Phys. Lett. \textbf{100A}(1984)467.\\

\par [8]V. Belinsky and V. E. Zakharov Sov. Phys.JETP \textbf{48(6)}1978.\\
\par [9] Jonathan J. C. Nimmo and Halis Yilmaz- On Darboux transformation for the derivative nonlinear Schr\"{o}dinger equation.arXiv1401.1536 v1[nonlin.SI]7 Jan 2014

\end{document}